\documentclass[aps,pre,reprint,showpacs]{revtex4-1}

\usepackage{graphicx}
\usepackage{amsmath, amsthm, amssymb}
\usepackage{textcomp}

\begin{document}

\title{Edge states of periodically kicked quantum rotors}

\author{Johannes Flo\ss}
\author{Ilya Sh.~Averbukh}
\affiliation{Department of Chemical Physics, Weizmann Institute of Science, 234 Herzl Street, Rehovot 76100, Israel}
\date{\today}

\begin{abstract}
We present a quantum localization phenomenon that exists in periodically kicked 3D rotors, but is absent in the commonly studied 2D ones: edge localization.
We show that under the condition of a fractional quantum resonance there are states of the kicked rotor that are strongly localized near the edge of the angular momentum space at $J=0$.
These states are analogs of surface states in crystalline solids, and they significantly affect resonant excitation of molecular rotation by laser pulse trains.
\end{abstract}

\pacs{05.45.-a, 37.10.Vz, 33.80.-b, 42.65.Re}


\maketitle


\section{Introduction}

The periodically kicked 2D rotor has been the subject of intensive research in the last four decades.
It is a standard model in studies on non-linear dynamics and quantum chaos~\cite{lichtenberg92,casati06,haake10}.
In the classical regime, a periodically kicked rotor can exhibit chaotic motion, leading to an unbounded growth of the angular momentum.
A quantum mechanical rotor shows chaotic-like behavior only for a limited period of time.
Eventually, the discreteness of the rotor's energy spectrum leads to at least quasiperiodic motion and therefore to a suppression of the diffusive growth of the angular momentum~\cite{casati79,fishman96}.
It was shown~\cite{fishman82,*grempel84} that this quantum suppression is due to a mechanism closely related to the Anderson localization of electronic wave functions in disordered solids~\cite{anderson58}.
Another distinct feature of the quantum kicked rotor is the  quantum resonance effect~\cite{casati79,izrailev80}.
A kick creates a rotational wave packet that revives after the so-called rotational revival time $t_{\mathrm{rev}}$~\cite{averbukh89,robinett04}, which is determined solely by the moment of inertia of the rotor.
Several kicks separated in time by an integer multiple of $t_{\mathrm{rev}}$ add constructively their action, and the angular momentum of the rotor grows ballistically (linearly) with the number of kicks.
The quantum resonance persists in a weakened form if the time-delay is equal to a rational multiple of $t_{\mathrm{rev}}$;
this effect is called a fractional quantum resonance.

The recent years saw a growing number of experiments utilizing the quantum resonance effect in linear molecules kicked by periodic trains of laser pulses.
Cryan~\textit{et al.} showed that a train of eight short laser pulses separated by the rotational revival time of molecular nitrogen leads to strong alignment of the molecules under standard conditions~\cite{cryan09}.
Other groups applied the quantum resonance effect for isotope-selective excitation~\cite{zhdanovich12a,akagi12a}, impulsive gas heating for Raman photoacoustics~\cite{schippers13} and controlling high power optical pulse propagation in open air~\cite{zahedpour14}.
In light of these experiments, a better understanding of the dynamics of the periodically kicked three-dimensional quantum rotor is desirable.

The periodically kicked 2D (planar) rotor, with one angular degree of freedom, has been intensively studied over the last 40 years.
Its sibling, the 3D (linear) rotor -- with two angular degrees of freedom, the polar angle $\theta$ and the azimuthal angle $\phi$ -- was only considered in a handful of studies (see~\cite{bluemel86,gong01,floss12,floss13}, and references therein), which furthermore concentrated on the similarities between the 2D and 3D cases.
Yet, there are qualitative differences between the two rotors.
The most obvious one is the edge in the angular momentum space:
Whilst for the planar rotor the angular momentum $J$ is unbounded ($-\infty<J<\infty$), for the 3D rotor no negative $J$ is allowed, thus there is an edge at $J=0$.
Matrix elements for the rotor coupling to the kicks are constant ($J$-independent) in 2D, and they take almost the same values in the three-dimensional case for large enough $J$.
However, the coupling becomes $J$-dependent near the edge $J=0$ in the 3D case.
Furthermore, the 3D rotor has an additional quantum number, the projection $M_J$ of the angular momentum on a space-fixed axis.

In this work, we present a remarkable phenomenon that exists in periodically kicked 3D rotors, but is absent in the commonly studied 2D ones: edge localization.
We numerically explore the 3D rotor excited at a fractional resonance, i.e. for a kicking period $\tau=(p/q)t_{\mathrm{rev}}$ ($p$ and $q$ being mutually prime).
We show that there are quantum states of the kicked rotor that are strongly localized near the edge of the angular momentum space at $J=0$.
As a result, if the initial state of a rotor lies near the edge, a major part of the population keeps being close to the edge regardless of the number of kicks applied.
Despite the fact that this phenomenon has such a dramatic effect on rotational excitation, it went practically unnoticed (except for an insightful hint in~\cite{bluemel86}) and has remained unexplored until now.

It was shown in the past~\cite{fishman82,grempel84,shepelyansky86} that the periodically kicked rotor can be mapped onto a tight-binding model known in solid state physics.
The spatial dimension in this model is the angular momentum $J$ (i.e. the levels $J$ are represented by the discrete grid sites of the model), and the coupling between the sites is due to the  kicks.
The edge of the momentum space at $J=0$ becomes an edge in the spatial grid of the tight-binding model, similar to a surface of a crystal.
It is known that the  surfaces in crystals can give rise to localized electronic states~\cite{ashcroft76}.
The edge states found near $J=0$ can be seen as the kicked rotor analogs of the surface states in  crystals.

This work is structured as follows.
In Sec.~\ref{sec.calculation} we introduce the model and numerical methods.
The main part of this work is Sec.~\ref{sec.results}, where we present our results for the quantum resonance in a linear rotor.
In particular, we show that for a fractional resonance one can find discrete quasienergy states that are localized at the edge of the angular momentum space.
We then show how these edge states influence the rotational dynamics.
In the last part of this section we consider special cases, like the quantum anti-resonance ($p/q=1/2$) and high-order resonances (large $q$).
In Sec.~\ref{sec.molecules}, we discuss the connection of our findings to current laser schemes for control of molecular rotations.
We also propose an experiment for observation of the edge localization.


\section{\label{sec.calculation}Model and numerical method}

We consider a rigid 3D rotor being periodically kicked by $\delta$-kicks.
In particular, we investigate the model of a linear rotor, described by two angular variables, the polar angle $\theta$ and the azimuthal angle $\phi$.
This model corresponds to, e.g., linear molecules like N$_2$, CO$_2$ or ICl interacting with a train of short laser pulses.

In this work, energy is given in units of $\hbar^2/I$ (where $I$ is the moment of inertia), and time in units of $I/\hbar$.
The rotational levels are $E_J=(1/2) J (J+1)$, where $J$ is the angular momentum quantum number.

The Hamiltonian for the system is given as
\begin{equation}
H=\frac{\hat J^2}{2} - P\cos^2\theta\sum_{n=1}^N \delta\left[t-(n-1/2)\tau\right] \,.
\label{eq.hamiltonian}
\end{equation}
Here, $\hat J$ is the angular momentum operator, $P$ is the strength of the kicks, $\theta$ is the polar angle, $N$ is the number of kicks, and $\tau$ is the periodicity of the kicks.
It should be noted that this interaction couples only angular momentum states of the same parity, $\Delta J =0,\pm2$.
Also, the projection $M_J$ of the angular momentum on the space-fixed $Z$-axis is conserved, so $M_J$ is a mere parameter defined by the initial conditions.

Different to most earlier studies on the kicked rotor, we use a $\cos^2\theta$ interaction instead of the common $\cos\theta$.
We chose this interaction potential having in mind experiments on laser control of molecular rotation.
The kick strength $P$ is related to experimental parameters via $P=(\Delta\alpha/4\hbar)\int \mathcal{E}^2(t) \mathrm{d}t$, where $\Delta\alpha$ is the molecular polarizability anisotropy and $\mathcal{E}(t)$ is the envelope of the electric field of the laser pulses.
Typically, the kick strength in current experiments is in the range of $1\lesssim P \lesssim 20$.

Since the Hamiltonian is that of a free rotor apart from the instant of the kick, it is helpful to expand the wave function of the rotor in the basis of the eigenfunctions of $\hat J^2$, the spherical harmonics $|J,M_J\rangle$:
\begin{equation}
|\Psi(t)\rangle = \sum_J C_J(t) e^{-iE_Jt} |J,M_J\rangle \,.
\label{eq.dynamics_sh}
\end{equation}
The expansion coefficients $C_J(t)$ are time-independent between the kicks, but change during a kick.
The structure of the rotational levels $E_J$ leads to exact revivals of any rotational wave packet after multiples of the  revival time $t_{\mathrm{rev}}=2\pi$.

A good way to understand the dynamics of a periodically driven quantum system is by looking at its quasienergy states (Floquet states)~\cite{zeldovich67}, the eigenstates of a one-cycle (pulse-to-pulse) evolution operator.
The quasienergy eigenstate $|\chi_{\alpha} \rangle (t)$ reproduces itself after a one-period evolution up to a certain phase factor, the quasienergy $\omega_{\alpha}$:
\begin{equation}\label{qes}
|\chi_{\alpha} \rangle (t+\tau)=e^{-i\omega_{\alpha}}|\chi_{\alpha} \rangle (t) .
\end{equation}
The quasienergy states can therefore be expressed as $|\chi_{\alpha}(t)\rangle=\exp(-i\omega_{\alpha}t/\tau)|u_{\alpha}(t)\rangle$, where $|u_{\alpha}(t+\tau)\rangle=|u_{\alpha}(t)\rangle$ is a time-periodic function.
Note that the value of the quasienergy is  defined only up to $\mod(2\pi)$.
By choosing the specific $2\pi$~interval, e.g. $-\pi\leq\omega_{\alpha}<\pi$, one uniquely defines $|u_{\alpha}(t)\rangle$.
One can represent the wave function of a periodically driven system as a linear combination of the quasienergy states~\cite{zeldovich67}:
\begin{equation}
|\Psi(t)\rangle = \sum_{\alpha} C_{\alpha} e^{-i\omega_{\alpha} t/\tau} |u_{\alpha}(t)\rangle \,.
\label{eq.dynamics_qe}
\end{equation}
The advantage of the expansion~\eqref{eq.dynamics_qe} is that the coefficients $C_{\alpha}$ are time-independent and are defined  by the initial state, $C_{\alpha}=\langle u_{\alpha}(0)|\Psi(0)\rangle$.
Therefore, the overlap of the initial state with the quasienergy states fully describes the time-dependent dynamics of the system.

For the numerical calculation, we directly solve the time-dependent Schr\"odinger equation with the Hamiltonian~\eqref{eq.hamiltonian}, using the spherical harmonics as a basis set, as described in detail in~\cite{fleischer09,floss12b}.
Thereby we obtain the  coefficients for the expansion in spherical harmonics, Eq.~\eqref{eq.dynamics_sh}.
In order to obtain the  coefficients for the expansion in quasienergy states, Eq.~\eqref{eq.dynamics_qe}, we first calculate the one-cycle evolution operator,
\begin{equation}
\hat U= e^{-i\hat J^2 \tau/4} e^{iP\cos^2\theta} e^{-i\hat J^2 \tau/4} \,.
\label{eq.evolutionoperator}
\end{equation}
The first and last term on the right-hand side of Eq.~\ref{eq.evolutionoperator} account for the free evolution before and after the kick, and the middle term accounts for the instantaneous kick.
The matrix elements of $\hat U$ in the basis of the spherical harmonics are obtained by solving the time-dependent Schr\"odinger equation for one cycle.
In particular, the element $U_{J',J}=\langle J',M_J|\hat U |J,M_J\rangle$ is given as $C_{J'}(\tau)e^{-iE_{J'}\tau}$ with the initial conditions $|\Psi(0)\rangle=|J,M_J\rangle$.
We numerically diagonalize $\mathbf{U}$ and thus obtain the quasienergies and the quasienergy states in the basis of the spherical harmonics.
Note that by this method we only obtain the quasienergy states at the start of each cycle, $|\chi_{\alpha}(0)\rangle=|u_{\alpha}(0)\rangle$.
This is sufficient to determine the expansion coefficients $C_{\alpha}=\langle u_{\alpha}(0)|\Psi(0)\rangle$.

The largest angular momentum $J_{\mathrm{max}}$ taken into account for the numerical simulations presented in this article is $J_{\mathrm{max}}=512$.
The lower bound is $J_{\mathrm{min}}=|M_J|$.
Note that even and odd $J$ form two independent subspaces (the interaction couples only states of the same parity).
Since we found no qualitative differences, in Sec.~\ref{sec.results} only the results for the even states are shown.
The artificial upper bound $J_{\mathrm{max}}$ can cause numerical artefacts.
In the presentation of the results we therefore exclude all states that are localized at this artificial upper bound~\footnote{We verified that these states are indeed numerical artefacts by varying the grid size.}.

For classification purposes, we consider a quasienergy state  as an edge state, if its overlap with the angular momentum states at the lower edge is at least 10\%, and its overlap at the upper (purely numerical) edge is less than $10^{-6}$,
\begin{align}
\sum_{J=J_{\mathrm{min}}}^{J_{\mathrm{min}}+39} |\langle \chi_{\alpha}(0)|J,M_J\rangle|^2 &>0.1 \nonumber\\
\sum_{J=J_{\mathrm{max}}-39}^{J_{\mathrm{max}}} |\langle \chi_{\alpha}(0)|J,M_J\rangle|^2 &< 10^{-6} \,.
\label{eq.edgedef}
\end{align}


\section{\label{sec.results}Results}

In this section we present the results of our numerical studies.
We first provide a short review of the quantum resonance effect in a 2D rotor.
We then present the quasienergy states as well as the quasienergy spectra for the quantum resonance in the 3D rotor.
We found that under the condition of a fractional resonance, most states form bands in the quasienergy spectrum and are delocalized in angular momentum space; however, some states are localized at the $J=0$ edge of the angular momentum space.
These states are found only at the edge of the quasienergy bands, or even completely remote from the bands.
In the third part of this section we show how these edge states manifest themselves in the rotational dynamics.
The fourth part is devoted to special cases, in particular the full and the half resonance, as well as high-order resonances.
In the last part we investigate the dependence of the results on the projection quantum number $M_J$, which is conserved in the the interaction;
we found that an increase of $|M_J|$ increases the edge effect, but there are no qualitative changes.
Apart from the last section, we only show the results for $M_J=0$.
We also only present the results for the states of even parity, as there is no qualitative difference between the two parities.

\subsection{The quantum resonance effect in a 2D rotor}

The rotational levels of a rigid 2D (planar) rotor are given as $E_J=J^2/2$.
Such a spectrum allows for quantum mechanical revivals:
Any wave packet of a 2D rotor revives exactly after integer multiples of the revival time, $t_{\mathrm{rev}}=4\pi$ (note the different revival time compared to the 3D rotor).
At rational multiples of $t_{\mathrm{rev}}$, so-called fractional revivals can be observed.
The revivals give rise to the quantum resonance effect:
Short kicks separated in time by rational multiples of the revival time, $\tau=(p/q)t_{\mathrm{rev}}$, add constructively their actions, and therefore the molecular angular momentum grows ballistically (linearly) with the number of pulses.

An early analysis of the quantum resonance in the planar rotor was provided by Izrailev and Shepelyanski~\cite{izrailev80}.
They showed analytically that over long times, the energy increases quadratically with the number of pulses.
Furthermore, the quasienergy spectrum consists of $q$ bands.
There can also be up to $q-1$ discrete levels, although we are not aware of any study that found those discrete levels.
For interaction strengths small compared to the order $q$ of the resonance, the bands are exponentially narrow.
A special case is the second order resonance $p/q=1/2$, also called quantum anti-resonance:
For this case, the quasienergy spectrum consists of only two values which differ by $\pi$.
Therefore, after two kicks separated by half the revival time any rotor returns exactly to its initial state.

\subsection{Quasienergy states and spectra}

\begin{figure}
\includegraphics[width=\linewidth]{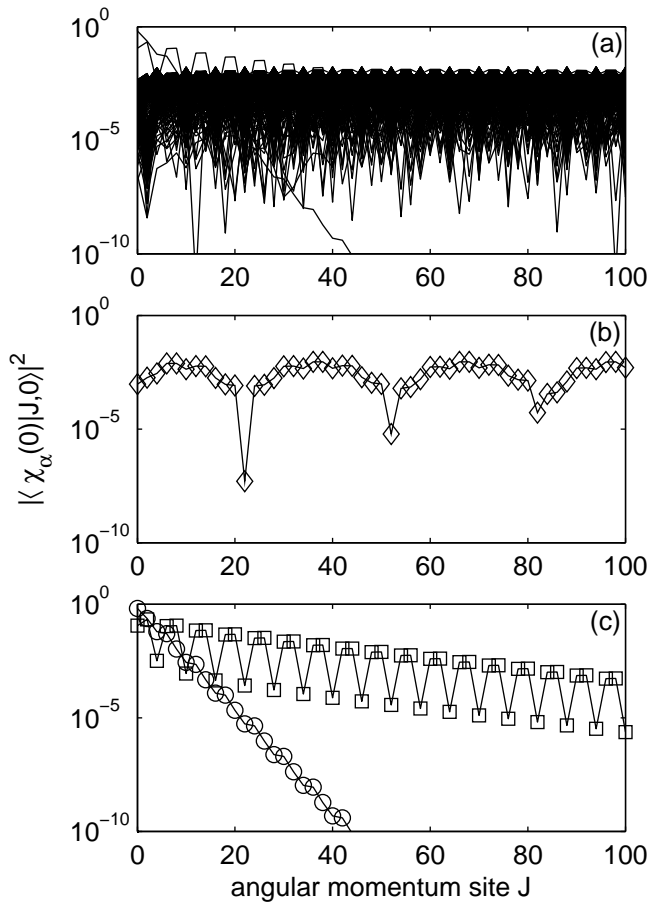}
\caption{
\label{fig.states_3}
Projection of the quasienergy states on the angular momentum states, for the case of the third order resonance $\tau=t_{\mathrm{rev}}/3$ and an interaction strength of $P=3$.
Panel~(a) displays the calculated quasienergy states; almost all states are extended over the full angular momentum space.
Panel~(b) shows one of the extended states separately, as an example.
In panel~(a), one can see two states that are localized at the lower edge of the momentum space; these states are shown separately in panel~(c).
Only states of even parity are shown.
}
\end{figure}

The results described here are generic for all fractional resonances that we have investigated.
For clarity, we will concentrate on one example, the third order resonance with a kicking period of $\tau=t_{\mathrm{rev}}/3$.
In Fig.~\ref{fig.states_3}~(a), the quasienergy states for this resonance for kicks of the strength of $P=3$ are presented.
In particular, the projection $|\langle \chi_{\alpha}(0)|J,0\rangle|^2$ of the quasienergy states on the angular momentum states is shown.
It can be seen that almost all states are extended over the whole angular momentum space.
For clarity, one of these states is shown separately in Fig.~\ref{fig.states_3}~(b); the other extended states look similar.
There are two states that are not extended, and are best described as edge states:
They have a maximum at $J=0$, and an exponentially decaying amplitude for increasing $J$.
These two edge states are also shown separately in Fig.~\ref{fig.states_3}~(c).
Such edge states can be found for most values of $P$.
Their number and localization lengths (``decay rate'') depend non-trivially on the kick strength $P$.

As can be seen in Fig.~\ref{fig.states_3}, the overlap of the edge states with the lowest rotational states can be quite large.
For the shown example, the rotational ground state $|0,0\rangle$ has an overlap of 75\% with the edge states:
The edge states dominate the dynamics of a system that is initially in (or close to) its ground state.
This is investigated in more detail below.

\begin{figure}
\includegraphics[width=\linewidth]{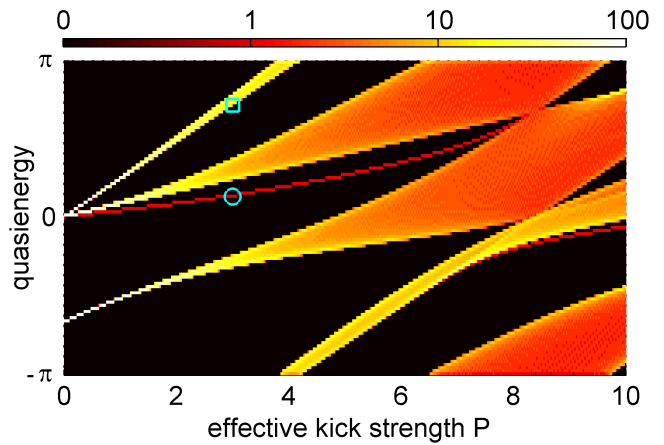}
\caption{
\label{fig.spectrum_3}
Spectrum of the quasienergy states for a rigid linear rotor kicked periodically at the fractional resonance $\tau=t_{\mathrm{rev}}/3$, as a function of the kick strength $P$.
The markers correspond to the states shown in Fig.~\ref{fig.states_3}~(c).
Only states of even parity are included.
The color axis depicts the numerical density, in particular the number of states per pixel; note its logarithmic scale.
}
\end{figure}

We now look at the quasienergy spectrum.
In Fig.~\ref{fig.spectrum_3}, we show the spectrum for the example of $p/q=1/3$, as a function of the kick strength $P$.
The color coding depicts the density of states.
One can clearly see three bands that broaden with the kick strength and eventually intersect.
Additionally, one can see two discrete levels.
One with a quasienergy between 0 and $\pi/2$, existing for $0<P<8.5$, and a second with a quasienergy of approximately $\pi/10$, emerging at $P\approx 7$.
These discrete states are localized on the $J=0$ edge, like the ones shown in Fig.~\ref{fig.states_3}~(c).
We also marked the position of the edge states from Fig.~\ref{fig.states_3}~(c) in the spectrum:
One is an discrete state, the other is at the edge of a band.

In our simulations, we observed the following patterns:
For a fractional resonance $\tau=(p/q) t_{\mathrm{rev}}$, the quasienergy spectrum consists of up to $q$ bands which broaden with increasing kick strength.
Additionally, discrete states exist for most interaction strength values .
These states are always localized at the $J=0$ edge, and vice versa, edge states are found only as discrete states or the states at the edge of a band.
We could not determine a definite rule for the number of edge states;
however, it seems that it increases with the order $q$ as well as the projection quantum number $|M_J|$ (see also below).


\subsection{\label{sec.dynamics}Dynamics}

\begin{figure}
\includegraphics[width=\linewidth]{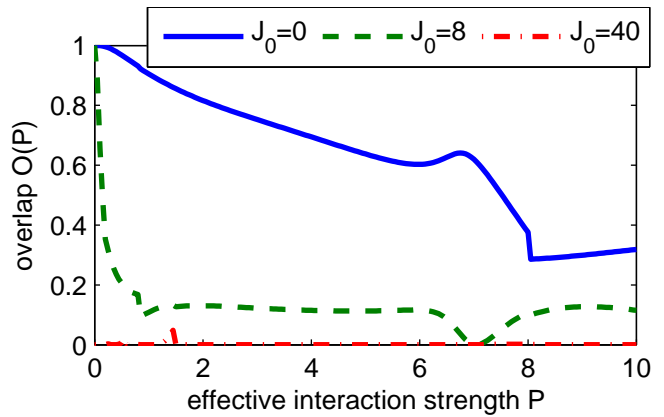}
\caption{
\label{fig.overlap_3}
Overlap $O(P)$ [see Eq.~\eqref{eq.overlap}] of different initial states $|\Psi(0)\rangle=|J_0,0\rangle$ with edge states as a function of the effective interaction strength $P$.
Shown is the case for the third order resonance $\tau=t_{\mathrm{rev}}/3$.
}
\end{figure}

Looking at Eq.~\eqref{eq.dynamics_qe}, we can see that the influence of the edge states on the dynamics can be quantified by the total overlap of the edge states with the initial state.
The overlap is given as
\begin{equation}
O(|\Psi (0)\rangle,P)= \sum_{\substack{\text{edge}\\\text{states}}} |\langle\chi_{\alpha}(t=0;P)|\Psi (0)\rangle|^2 \,,
\label{eq.overlap}
\end{equation}
where the sum is over all edge states (as defined in Sec.~\ref{sec.calculation}).
Intuitively, we would expect that an initial state lying further away from the edge has less overlap with the edge states.
Also, an increase of the kick strength $P$ should decrease the influence of the edge states, since a stronger kick couples angular momentum states at the edge more effectively with states further away from the edge.
Our numerical results support these intuitive guesses.
However, the dependence on $P$ is non-trivial so one may find special values of $P$ for which the overlap is very large or very low.
In Fig.~\ref{fig.overlap_3} we show the overlap $O(P)$ for the above example of $\tau=t_{\mathrm{rev}}/3$, for three initial states $|0,0\rangle$, $|8,0\rangle$ and $|40,0\rangle$.
These initial states qualitatively represent the cases of a cold molecule, a nitrogen molecule at room temperature,  and a nitrogen super-rotor~\cite{korobenko13}, respectively.
For the cold molecule we find a large overlap of the rotational wave function with the edge states.
It decreases with increasing interaction strength, but even at $P=10$ still more than 30\% of the initial population is trapped in the localized edge state.
The decrease of the overlap with $P$ is, generally, monotonic, but also shows some local extrema;
in the shown example around $P=7$.
The jump seen at $P\approx8$ is a result of our convention in defining the edge states [see Eqs.~\eqref{eq.edgedef}], and it results from a single edge state turning into an extended one at this specific $P$-value.
For a typical nitrogen thermal state $|8,0\rangle$ taken as initial state, the edge states dominate the dynamics only for weak kicks of $P<1$.
For stronger kicks they have a minor influence and contribute only about 10\%.
Mind that for a thermal molecular rotor, also states with $M_J\neq0$ -- for which the edge localization is stronger (see Sec.~\ref{sec.mj}) -- are populated.
Finally, for the fast spinning initial state, there is only a marginal overlap with the edge states.

From these findings we  conclude that for a rotor in a low-lying initial state, the dynamics are dominated by the edge states, and  a significant part  of the population remains trapped close to the $J=0$ edge.
However, the untrapped part of the population belongs to the extended quasienergy states and, therefore, undergoes the quantum resonant excitation.
We can  expect the rotational energy to grow quadratically with the number of pulses even for a rotor initially in the ground state, although the growth is significantly reduced compared to the regular quantum resonance due to the edge effects.
For a rotor in a fast spinning initial state with no overlap with the edge states, there is no edge effect and the quantum resonance is unhindered.

\begin{figure}
\includegraphics[width=\linewidth]{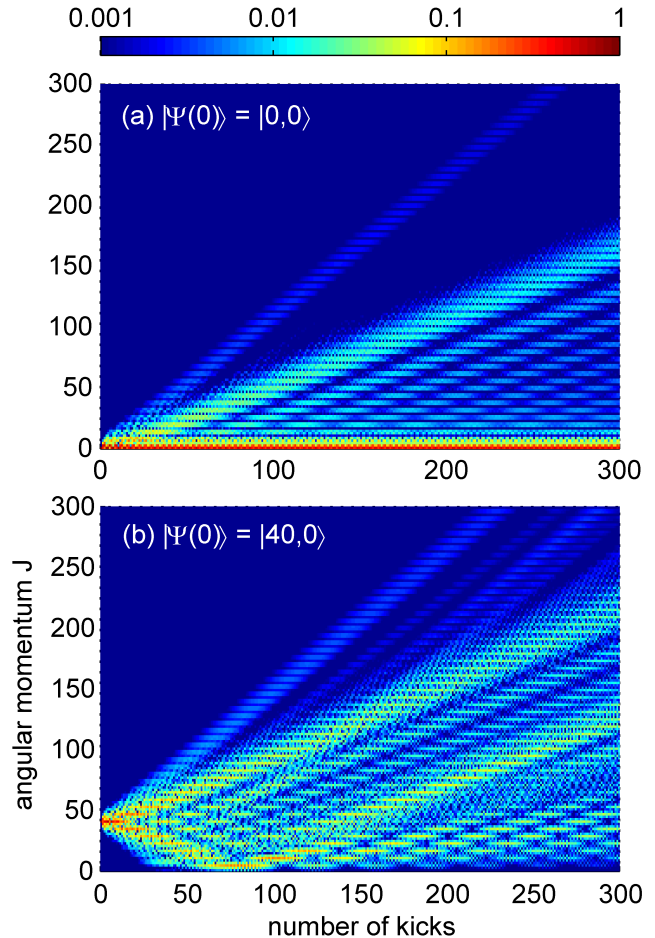}
\caption{
\label{fig.population_3}
Time-dependent population of the angular momentum states $|J,0\rangle$ for a rotor kicked periodically at the fractional resonance $\tau=t_{\mathrm{rev}}/3$ with kick strength $P=3$.
The initial state is (a) $|\Psi(0)\rangle=|0,0\rangle$ and (b) $|\Psi(0)\rangle=|40,0\rangle$.
Shown are only the states of even parity.
Note the logarithmic scale of the color axis.
}
\end{figure}

\begin{figure}
\includegraphics[width=\linewidth]{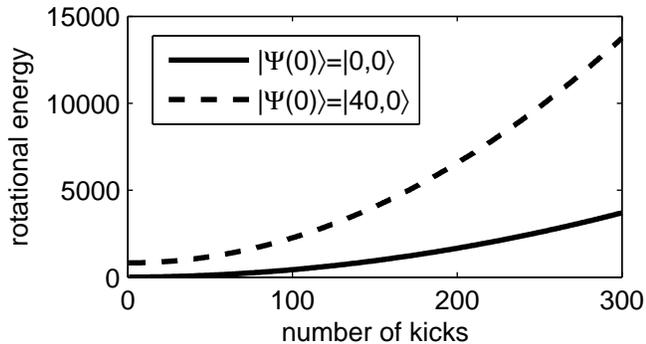}
\caption{\label{fig.energy_3}
Rotational energy as a function of the number of kicks, for a rotor initially in its ground state $|0,0\rangle$ (solid line) and an excited state $|40,0\rangle$ (dashed line).
The kicking period is $\tau=t_{\mathrm{rev}}/3$, the kick strength $P=3$.
}
\end{figure}

To demonstrate these conclusions, we show in Fig.~\ref{fig.population_3} the population of the angular momentum states as a function of the number of kicks, for  $\tau=t_{\mathrm{rev}}/3$ and $P=3$.
For an initial state with a low angular momentum [presented in Fig.~\ref{fig.population_3}~(a) with $|\Psi(0)\rangle=|0,0\rangle$], the dynamics are divided into a localized edge part and a delocalized resonant part:
a large fraction of the population keeps being close to the lower edge, whilst the remainder shows an unbounded linear growth of the angular momentum, the signature of the quantum resonance.
For a high-lying initial state [Fig.~\ref{fig.population_3}~(b), $|\Psi(0)\rangle=|40,0\rangle$], one can observe two streams of rotational excitation, directed towards higher and lower $J$.
The physical reason for the double stream is that in half of the angular space the direction of the kick coincides with the initial rotational velocity of the rotor, whilst in the other half it is directed oppositely.
When the downward stream reaches the $J=0$ edge, it is reflected, and over a sufficiently large number of kicks no population remains close to the edge.
The edge states have no effect on the dynamics of the rotors with a large initial angular momentum.
Also the rotational energy, displayed in Fig.~\ref{fig.energy_3}, shows the predicted behavior.
Both for the low-lying and the high-lying initial state the rotor energy grows quadratically with the number of pulses;
however, for the low-lying initial state the growth is much slower.

\subsection{\label{sec.specialcases}Special cases}

\begin{figure}
\includegraphics[width=\linewidth]{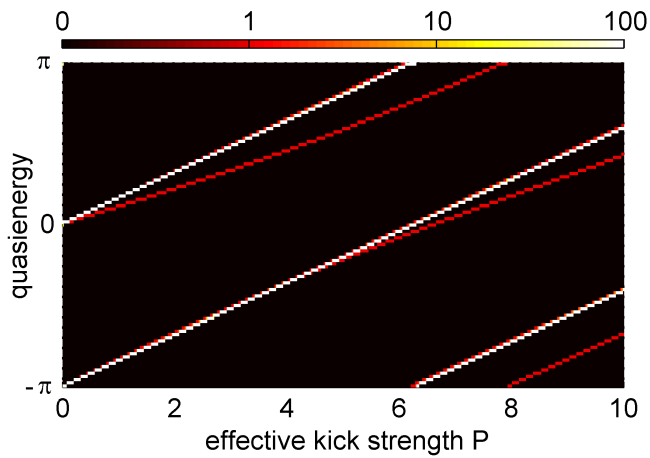}
\caption{
\label{fig.spectrum_2}
Density of quasienergy states for a rigid linear rotor kicked periodically at the fractional resonance $\tau=t_{\mathrm{rev}}/2$, as a function of the kick strength $P$.
Only states of even parity are included.
The color axis depicts the numerical density, in particular the number of states per pixel; note its logarithmic scale.
}
\end{figure}

A special case of the quantum resonance for a 2D rotor is the second order resonance at $\tau=t_{\mathrm{rev}}/2$.
This fractional resonance is also called anti-resonance: Instead of strong excitation, every second kick destroys the effect of the preceding one~\cite{izrailev80}.
All quasienergy states are degenerate, and there are only two possible values for the quasienergy which differ by exactly $\pi$.
Using Eq.~\eqref{eq.dynamics_qe}, one can easily see that such a spectrum leads to an exact revival of the initial state after two pulses.

\begin{figure}
\includegraphics[width=\linewidth]{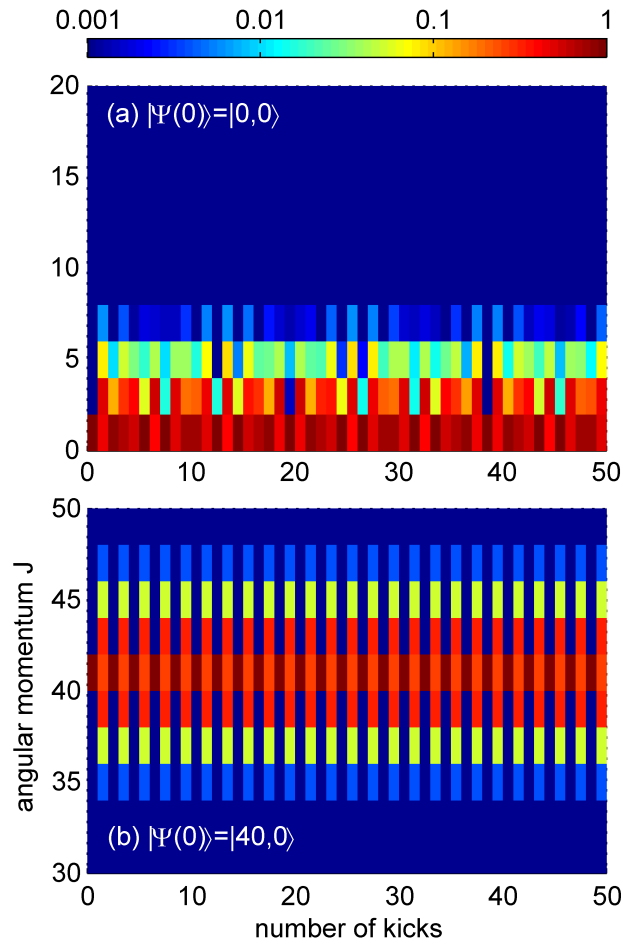}
\caption{
\label{fig.population_2}
Time-dependent population of the angular momentum states $|J,0\rangle$ for a rotor kicked periodically at the fractional resonance $\tau=t_{\mathrm{rev}}/2$ with kick strength $P=3$.
The initial state is (a) $|\Psi(0)\rangle=|0,0\rangle$ and (b) $|\Psi(0)\rangle=|40,0\rangle$.
Shown are only the states of even parity.
Note the logarithmic scale of the color axis.
}
\end{figure}

Also for the 3D rotor the second order resonance acts as an anti-resonance~\cite{lee06,fleischer06}.
However, this anti-resonance is not exact, and the rotational state only approximately revives after the second kick.
As we show in the following, this can be interpreted as an edge effect.
Looking at the quasienergy spectrum for $\tau=t_{\mathrm{rev}}/2$ (Fig.~\ref{fig.spectrum_2}), one can see two lines of (almost) degenerate states, with a difference of $\pi$.
Furthermore, there are discrete states;
our simulations show that these are edge states.
Since the edge states have a quasienergy which differs from the other quasienergies by a value different from $\pi$, there is no exact revival after two pulses.
Instead, one can see a quasi-revival at later times; e.g., for $P=3$, this quasi-revival happens after 38 pulses (see Fig.~\ref{fig.population_2}~(a),  showing the population of the angular momentum states as a function of the number of pulses).
Since this is an edge effect, we do not expect it to affect rotors with a high initial angular momentum.
Indeed, for fast rotating rotors, the exact anti-resonance is recovered, as shown in Fig.~\ref{fig.population_2}~(b):
After every other pulse the system returns to the initial state, here $|40,0\rangle$.
Note that for the interaction potential   considered in this work [see Eq.~\eqref{eq.hamiltonian}], also the fourth order resonances are anti-resonances: after four pulses, the system returns approximately to its initial state.

\begin{figure}
\includegraphics[width=\linewidth]{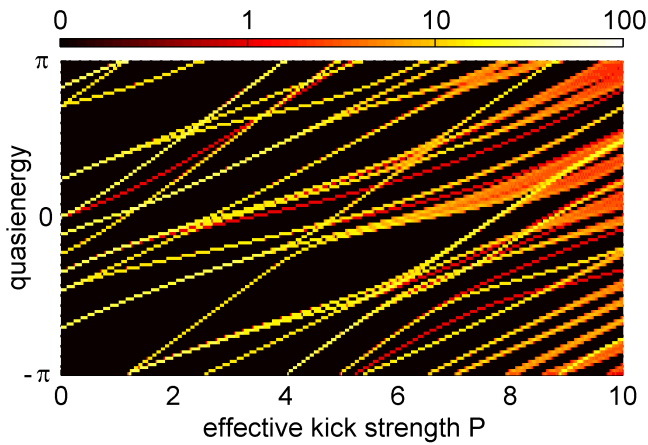}
\caption{
\label{fig.spectrum_17}
Density of quasienergy states for a rigid linear rotor kicked periodically at the high order fractional resonance $\tau=t_{\mathrm{rev}}/17$, as a function of the kick strength $P$.
Only states of even parity are included.
The color axis depicts the numerical density, in particular the number of states per pixel; note its logarithmic scale.
}
\end{figure}

Higher order resonances show a different behavior for low kick strengths $P$.
For a 2D rotor kicked at a resonance $\tau=(p/q)t_{\mathrm{rev}}$ with $q\gg1$, the quasienergy bands are exponentially narrow and effectively degenerate ~\cite{izrailev80}.
We found the same for the periodically kicked 3D rotor, as shown in Fig.~\ref{fig.spectrum_17} for $q=17$.
One can clearly see very narrow bands of almost degenerate states for $P<5$.
As for the quantum anti-resonance, this leads to localization instead of resonant excitation.
When $P$ becomes larger, the normal fractional resonance (as described above) is recovered.

The last special case we have to consider is the full resonance, $\tau=t_{\mathrm{rev}}$.
For this resonance, each component of the rotational wave packet accumulates a phase of an integer multiple of $2\pi$ in the course of every excitation period.
Thus, all kicks add their actions fully constructively, leading to fast developing rotational excitation.
Unlike fractional resonances, the full resonance does not show any edge states.


\subsection{\label{sec.mj}$M_J\neq0$}

\begin{figure}
\includegraphics[width=\linewidth]{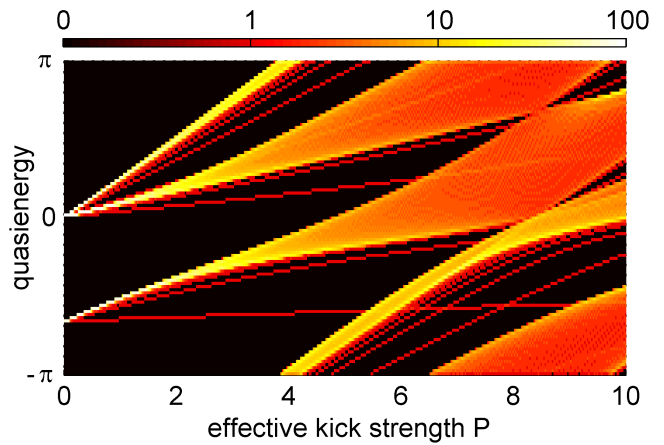}
\caption{
\label{fig.spectrum_3M10}
Density of quasienergy states for a rigid linear rotor kicked periodically at the fractional resonance $\tau=t_{\mathrm{rev}}/3$, as a function of the kick strength $P$.
The projection quantum number is $M_J=10$.
Only states of even parity are included.
The color axis depicts the numerical density, in particular the number of states per pixel; note its logarithmic scale.
}
\end{figure}

Up to now we only considered the case of $M_J=0$.
For $M_J\neq0$, our simulations show that all findings for the $M_J=0$ case are still valid.
There is one interesting difference:
With increasing $|M_J|$, also the number of edge states increases.
This can be seen in the quasienergy spectrum for $M_J=10$, shown in Fig.~\ref{fig.spectrum_3M10}.
There is a much larger number of discrete states than for the $M_J=0$ case (Fig.~\ref{fig.spectrum_3}), which further increases for larger values of $|M_J|$.
As before, these discrete states are localized at the edge, which is now at $J=|M_J|$.
We verified that the larger number of edge states is not a numerical artefact by repeating the calculation for different sizes of the $J$ grid.


\section{\label{sec.molecules}Edge localization and laser kicked molecules}

Experimental studies on the periodically kicked rotor have routinely been done on cold atoms interacting with a pulsed standing light wave~\cite{moore95,raizen99,klappauf98,amman98}, a system imitating the dynamics of the kicked 2D rotor.
Very recently a new kind of experiments has appeared, using linear molecules kicked by periodic trains of short laser pulses.
In these experiments, the quantum resonance~\cite{cryan09,floss12,zhdanovich12a}, Anderson localization~\cite{bluemel86,floss12,floss13,kamalov15}, and Bloch oscillations~\cite{floss14,floss15} have been observed.
The 3D localization phenomenon presented in this work may also be observed in laser kicked molecules.

\begin{figure}
\includegraphics[width=\linewidth]{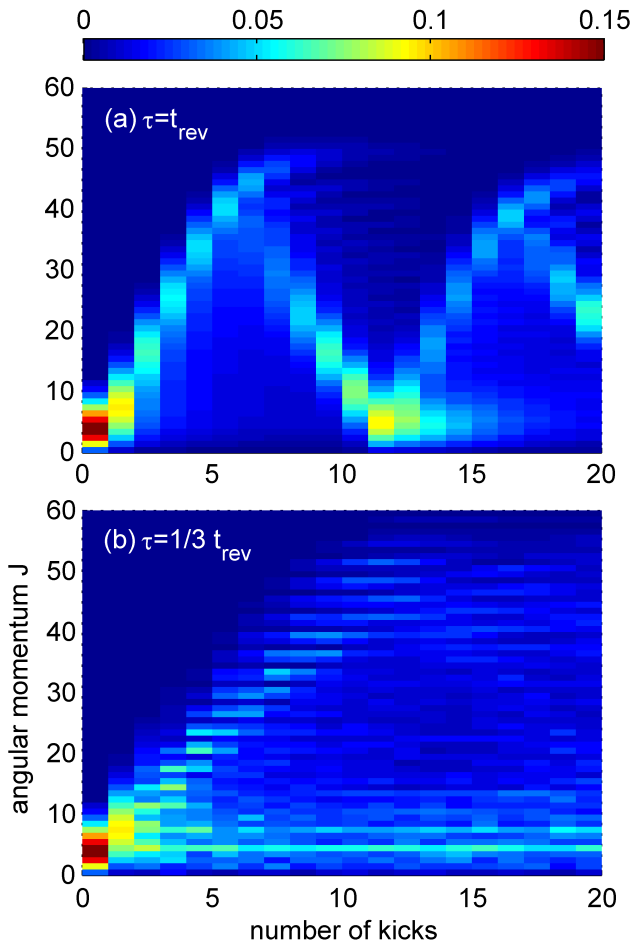}
\caption{\label{fig.population_icl}
Simulated population of the angular momentum levels $J$ for $^{129}$I$^{35}$Cl molecules kicked periodically at (a) the full resonance ($\tau=t_{\mathrm{rev}}=146.1~\text{ps}$), and (b) the third order resonance ($\tau=t_{\mathrm{rev}}/3$).
The pulse duration is 500~fs (full width at half maximum), the peak intensity is 1.5~TW/cm$^2$.
This corresponds to a kick strength of $P=10$.
The initial rotational temperature is set to 5~K.
}
\end{figure}

\begin{figure}
\includegraphics{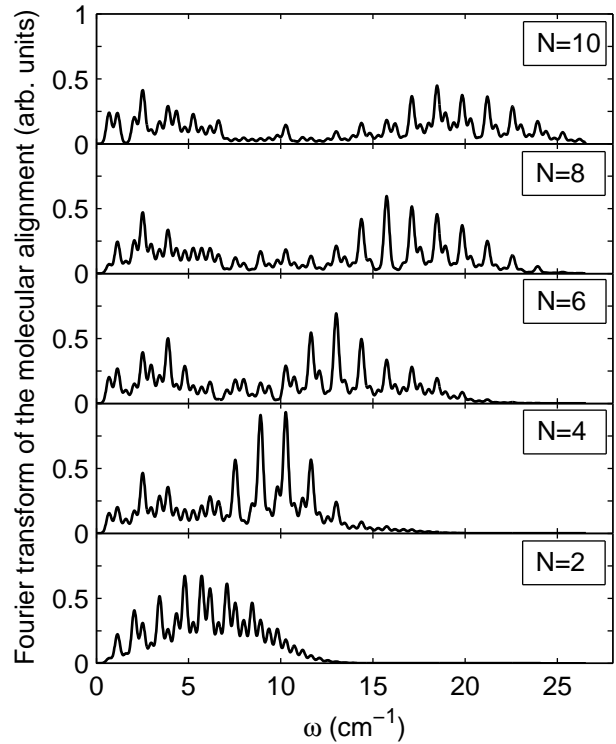}
\caption{\label{fig.FTalignment}
Absolute value of the Fourier transform of the molecular alignment signal $\langle \cos^2\theta \rangle (t)$ for ICl molecules kicked at $\tau=t_{\mathrm{rev}}/3$ [same conditions as for Fig.~\ref{fig.population_icl}~(b)].
The results are shown after $N$ pulses, for $N=2,4,6,8,10$.
The zero-frequency component (time-averaged alignment) is removed from the signal.
We added a broadening of 0.33~cm$^{-1}$, which corresponds to a measurement window of about 100~ps.
}
\end{figure}

To demonstrate this possibility, we simulate the rotational excitation of non-rigid ICl molecules kicked by a train of 20 pulses, using the numerical procedure described in~\cite{floss12b}.
As effective kick strength we choose $P=10$.
We include thermal effects by ensemble averaging over the initial states, where we choose 5~K as the initial temperature; we assume that collisions are negligible over the duration of the pulse train.
Such conditions can be achieved, e.g., in a molecular beam.
The results of the simulation are shown in Fig.~\ref{fig.population_icl}.
For a train tuned to the full resonance [Fig.~\ref{fig.population_icl}~(a)], almost all population shifts to higher and higher momentum states during the first pulses.
After about six pulses, the excitation is reversed, and subsequently oscillations of the population distribution are observed.
They are the rotational analog of Bloch oscillations and are caused by the non-rigidity of molecular rotors~\cite{floss14}.
For a train tuned to a fractional resonance ($\tau=t_{\mathrm{rev}}/3$) [Fig.~\ref{fig.population_icl}~(b)], one can see two streams of excitation.
The first one shows the expected quantum resonance behavior --
the angular momentum grows linearly with the number of pulses, up to $J\sim60$.
The second stream is the manifestation of the edge localization:
A large part of the population (about one third) remains close to the edge ($J\lesssim10$), regardless of the number of pulses applied.
This splitting of the population can be measured by direct methods, e.g. by resonance enhanced multiphoton ionization, as done in~\cite{zhdanovich12a}.
Alternatively, one can measure the time-dependence of the birefringence (caused by molecular alignment).
This signal will show a modulation with two clearly separated frequency groups:
high frequencies corresponding to the resonantly excited stream, and low frequencies caused by the edges states.
This is demonstrated in Fig.~\ref{fig.FTalignment}, where we show the Fourier transform of the molecular alignment $\langle \cos^2\theta \rangle (t)$ after $N=2,4,6,8$, and $10$ pulses.
One can clearly see the splitting of the Fourier components into a high-frequency part which shifts to higher frequencies with increasing $N$, and a low-frequency part which remains more or less unaltered irrespective of the number of pulses applied.

\section{\label{sec.discussion}Conclusion}

In this work, we presented the first thorough study of a quantum localization phenomenon that exists in the periodically kicked 3D rotor:
edge localization of the rotational excitation.
We showed by the help of numerical simulations that under the condition of the fractional quantum resonance -- when one may expect an unhindered rotational excitation --, there are quasienergy states localized near the edge of the angular momentum space at $J=0$.
These states lead to a trapping of a considerable part of the rotational population close to the edge.
The corresponding quasienergies are either discrete or found to be at the edge of a quasienergy band.
This effect is completely absent in the commonly studied 2D rotor.
These states can be considered as the rotational analog of the surface states in a crystalline solid~\cite{ashcroft76}.

The edge localization adds nicely to two other quantum localization phenomena in rotational systems which have analogs in solid state physics:
Anderson localization~\cite{fishman82} and Bloch oscillations~\cite{floss14}.
The latter two phenomena appear only for pulse trains detuned from the quantum resonance, whilst the effect presented in this work exists on resonance.

We showed that the edge localization can be observed in current experimental schemes that are used to explore the periodically kicked rotor, namely linear molecules interacting with periodic trains of short laser pulses.
Our work shows that laser excitation of molecular rotation can be strongly affected by the edge localization.

\begin{acknowledgements}
Financial support of this research by the ISF (Grant No. 601/10), the DFG (Project No. LE 2138/2-1), and the Minerva Foundation is gratefully acknowledged.
I.~A.~acknowledges support as the Patricia Elman Bildner Professorial Chair.
This research was made possible in part by the historic generosity of the Harold Perlman Family.
\end{acknowledgements}



%

\end{document}